\documentclass[aip,reprint]{revtex4-1}

\usepackage{amsmath}
\usepackage{color}
\usepackage{graphicx}
\usepackage{natbib}
\usepackage{units}

\begin{document}

\title{Ion beam synthesis of nanothermochromic diffraction gratings with giant switching contrast at telecom wavelengths}

\author{Johannes Zimmer}
\affiliation{Lehrstuhl f\"{u}r Experimentalphysik 1 and Augsburg Centre for Innovative Technologies (ACIT), Universit\"{a}t Augsburg, Universit\"{a}tstr. 1, 86159 Augsburg, Germany} 
\author{Achim Wixforth}
\affiliation{Lehrstuhl f\"{u}r Experimentalphysik 1 and Augsburg Centre for Innovative Technologies (ACIT), Universit\"{a}t Augsburg, Universit\"{a}tstr. 1, 86159 Augsburg, Germany} 
\author{Helmut Karl}
\affiliation{Lehrstuhl f\"{u}r Experimentalphysik IV, Universit\"{a}t Augsburg, Universit\"{a}tstr. 1, 86159 Augsburg, Germany} 
\author{Hubert J. Krenner}\email{hubert.krenner@physik.uni-augsburg.de}
\affiliation{Lehrstuhl f\"{u}r Experimentalphysik 1 and Augsburg Centre for Innovative Technologies (ACIT), Universit\"{a}t Augsburg, Universit\"{a}tstr. 1, 86159 Augsburg, Germany} 

\begin{abstract}
	Nanothermochromic diffraction gratings based on the metal-insulator transition of $\mathrm{VO_2}$ are fabricated by site-selective ion beam implantation in a $\mathrm{SiO_2}$ matrix. Gratings were defined either (i) directly by spatially selective ion beam synthesis or (ii) by site-selective deactivation of the phase transition by ion beam induced defects. The strongest increase of the diffracted light intensities was observed at a wavelength of 1550\,nm exceeding a factor of 20 for the selectively deactivated gratings. The observed pronounced thermal hysteresis extending down close to room temperature makes this system ideally suited for optical memory applications. 
\end{abstract}

\maketitle


Vanadium dioxide $(\mathrm{VO_2})$ has been studied for more than five decades in great detail since it exhibits a metal-insulator transition (MIT) which occurs in bulk crystals at $T_C\sim68\mathrm{^oC}$. At this MIT the material undergoes a structural and electronic phase transition which gives rise to a dramatic increase of the electrical conductivity \cite{Morin:59} when the material is heated up from the insulating phase at room temperature to the high temperature metallic phase. In addition to the DC electrical conductivity the complex index of refraction of $\mathrm{VO_2}$ changes at the MIT \cite{Verleur:68,Liu:11} which can moreover be controlled on sub-picosecond timescales \cite{Cavalleri:04a,Rini:05a}. This opens directions for fast switching of optical elements and electronic devices\cite{Yang:11} which operate close or slightly above room temperature. This thermochromicity is most pronounced in the near infrared down to the terahertz spectral domain and first steps have been taken towards tunable optical elements such as tunable absorptive coatings \cite{Lopez:04a} and fiber optical \cite{Lee:89} or planar photonic elements\cite{Briggs:10}. \\

Here we present two routes to define nanothermochromic $\mathrm{VO_2}$ diffractive optical devices in a $\mathrm{SiO_2}$ matrix using site-selective ion beam implantation. The first approach is based on direct site-selective ion beam synthesis of $\mathrm{VO_2}$ nanoclusters while in the second "cold" process the MIT in sub-ensembles of nanocrystals is inhibited by selective introduction of point defects using argon ion bombardment. To demonstrate the feasibility of these approaches we fabricated phase gratings which show a variation of diffraction efficiency of more than a factor of 3 for directly synthesized nanoclusters and more than one order of magnitude for gratings for which the MIT was selectively deactivated by ion beam induced defects. Moreover, this nanothermochromic switching is most pronounced at the technologically most relevant wavelength of 1550\,nm. Both types of gratings show a large thermal hysteresis which extends down close to room temperature when the system is cooled from the metallic state and is of high technological relevance for integrated optical storage devices.\\

\begin{figure}
	\begin{center}
		\includegraphics[width=0.75\columnwidth]{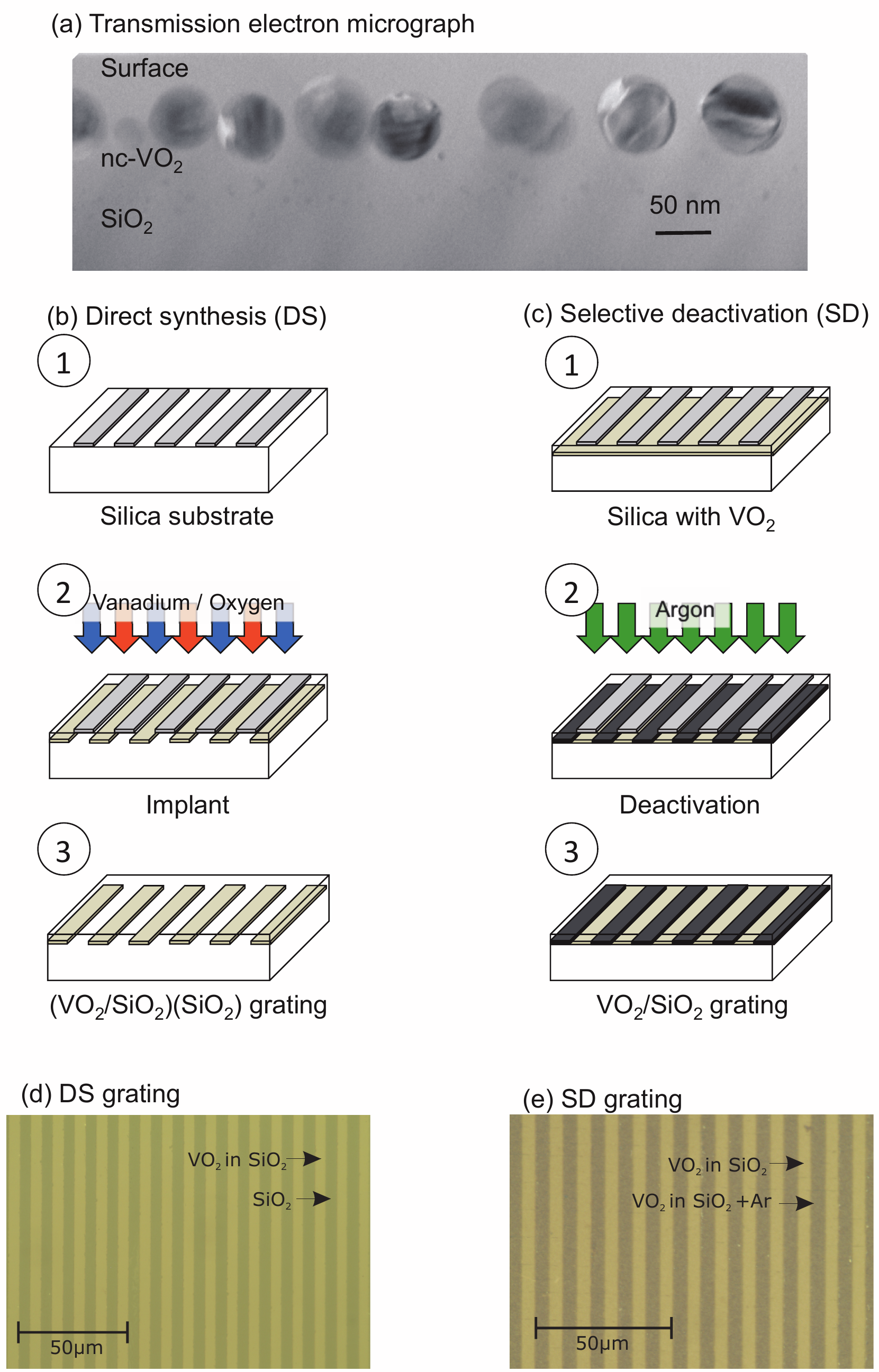}
		\caption{(Color online) (a) Cross-section transmission electron micrograph of $\mathrm{VO_2}$ nanocrystals embedded in a $\mathrm{SiO_2}$ matrix. Schematic of fabrication steps for DS gratings (b) and SD gratings (c). Microscope images of the studied DS (d) and SD gratings (e) with $\Lambda = 10~\mathrm{\mu m}$. and $DC=0.5$.}
		\label{fig:1}
	\end{center}
\end{figure}
The ion beam synthesis of $\mathrm{VO_2}$ nanocrystals\cite{Lopez:02a} starts with a two-step implantation process of vanadium ($9\cdot10^{16}\frac{at}{cm^2}$ @ 100\,keV) and oxygen ($1.8\cdot10^{17}\frac{at}{cm^2}$ @ 36\,keV) directly into a 0.5\,mm thick fused silica substrate for diffraction gratings. The $\mathrm{VO_2}$ nanocrystals are formed during a 10\,min rapid thermal annealing (RTA) step at 1000\textsuperscript{o}C. The such synthesized clusters have average diameters of 90\,nm and are centered 85\,nm below the sample surface as seen in the transmission electron micrograph presented in Fig. \ref{fig:1}(a). The two different routes to define nanothermochromic diffraction gratings are compared in Fig. \ref{fig:1}(b) and (c). For both approaches we employed 120\,nm thick chromium implantation masks defined by standard optical lithography in a lift-off process. This mask was removed after implantation using a selective wet chemical etch.\\

In the first approach shown in Fig. \ref{fig:1} (b) $\mathrm{VO_2}$ nanocrystals were synthesized by implantation of vanadium and oxygen through the implantation mask. In this bottom-up approach a RTA step is required after mask removal. We refer to the such fabricated gratings as \emph{directly synthesized} (DS) gratings. For the top-down process shown in panel (c) we start with a substrate containing a homogeneous layer of $\mathrm{VO_2}$ nanocrystals. In this approach the mask protects arrays of nanocrystals during an $\mathrm{Ar^+}$ ion implantation step ($7\cdot10^{15}\frac{at}{cm^2}$ @ 80\,keV). This energy was chosen to have the maximum energy loss per unit distance centered around 70\,nm close to the center of the $\mathrm{VO_2}$ nanocrystal layer. The bombardment with $\mathrm{Ar^+}$ introduces defects in the $\mathrm{VO_2}$ nanocrystals which inhibit their MIT. Due to this property we refer to these gratings as \emph{selectively deactivated} (SD) in the following. We want to note, that no further thermal treatment is required after mask removal. Thus, SD gratings can be readily defined in a "cold" process in contrast to their DS counterparts. Both types of two gratings are clearly resolved in the optical micrographs of Fig. \ref{fig:1} (d) and (e).\\
\begin{figure*}
	\begin{center}
		\includegraphics[width=1.95\columnwidth]{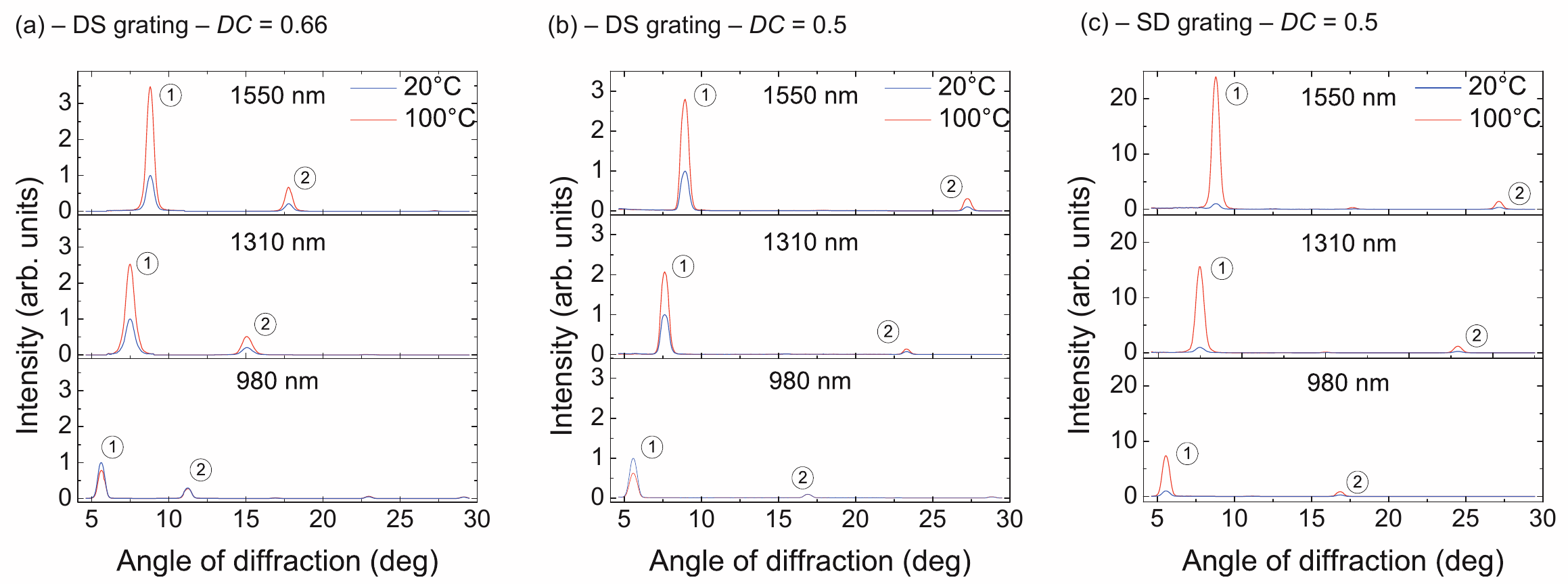}
		\caption{(Color online) Angle-resolved diffraction scan for DS gratings with $DC=0.66$ (a) and $DC=0.5$ (b) and a SD grating with $DC=0.5$ (c) showing two dominant diffraction peaks labeled 1 and 2 in the insulating ($T=20~\mathrm{^oC}$, blue lines) and the metallic phase ($T=100~\mathrm{^oC}$, red lines). Intensities are normalized to peak 1 in the insulating phase for each grating and wavelength.}
		\label{fig:2}
	\end{center}
\end{figure*}

We studied these gratings using angle-resolved light diffraction using commercial laser diodes at three technologically most relevant wavelengths of 980\,nm, 1310\,nm and 1550\,nm. The samples themselves were mounted on a temperature-controlled holder in the center of a double-stage goniometer (diameter 25 cm) which allows for independent tuning of the angle of incidence and the detector. All experimental data presented here were taken under normal incidence in transmission and the diffracted light was detected with an angular resolution of $<0.5$ degrees using a standard $\mathrm{InGaAs}$ photodiode.\\
We present angle-resolved diffraction scans for all three wavelengths for gratings with a periodicity $\Lambda=10~\mathrm{\mu m}$ in Fig. \ref{fig:2} at room temperature $(T=20~\mathrm{^oC}$, blue lines) and at elevated temperature $(T=100~\mathrm{^oC}$, red lines). At these temperatures the $\mathrm{VO_2}$ nanoclusters are in the insulating and metallic state, respectively. We compare two DS gratings with duty cycles $DC=0.66$ and $0.5$ of $(\mathrm{VO_2/SiO_2}):(\mathrm{SiO_2})$ in Fig. \ref{fig:2} (a) and (b) and present a corresponding diffraction scan of a SD grating with $DC=0.5$ in Fig. \ref{fig:2} (c). The intensities in each panel are normalized to the principal diffraction order labeled "1" in the insulating state. We find all diffraction angles in good agreement with the nominal $\Lambda$. Moreover, the next higher \emph{dominant} diffraction order ("2") shows a pronounced dependence on $DC$. As expected it clearly shifts towards smaller angles for the DS grating with $DC=0.66$ in Fig. \ref{fig:2} (a) compared to the DS and SD gratings in panels (b) and (c) with $DC=0.5$. \\

The most striking difference is observed for the variation of the diffraction efficiencies of the two grating types. Diffraction is pronounced for the DS grating due to the relatively large refractive index contrast between the non implanted $\mathrm{SiO_2}$ and the $\mathrm{VO_2/SiO_2}$ for both low and high temperatures. As $\mathrm{VO_2}$ undergoes the MIT into the metallic state the diffraction efficiency decreases for $\lambda=980\mathrm{~nm}$ and increases for both $\lambda=1310\mathrm{~nm}$ and $\lambda=1550\mathrm{~nm}$. This variation arises from an enhancement (reduction) of the refractive index contrast at the two longer (short) wavelength. Thus, for our $\mathrm{VO_2/SiO_2}$ composite the variation of the refractive index at the MIT changes sign in the near-infrared spectral domain which is also confirmed by spectral ellipsometry investigation on unstructured samples. In particular the increase of the refractive index is typically not observed in bulk and thin film samples \cite{Verleur:68,Briggs:10,Tazawa:98}. Therefore, we attribute this anomalous switching behavior to small cluster size. Moreover, we can exclude a dominant contribution due to plasmonic effects. These would result in additional loss in the metallic state\cite{Rini:05a,Li:10} and are only weakly wavelength dependent \cite{Verleur:68,Tazawa:98} in this spectral range.
In strong contrast, the SD grating diffraction is weak at room temperature due to the small refractive index contrast generated by the ion bombardment. As the \emph{active} $\mathrm{VO_2}$ nanocrystals undergo the MIT to the metallic state, their refractive index changes. The result in a giant variation of the dielectric contrast of the grating. This in turn gives rise to the observed giant enhancement of the diffraction efficiency for \emph{all} three wavelengths studied. \\

We further quantify this effect by evaluating the switching contrast $\tilde{I}_n=\frac{I_{m,n}}{I_{i,n}}$. Here $I_{m,n}$ and $I_{i,n}$ denote the intensities of the $n$-th diffraction order in the metallic $(m)$ and insulating $(i)$ state, respectively. The extracted values of this figure of merit are summarized in Tab. \ref{tab:1} for the dominant diffraction order $(n=1)$. Clearly, a pronounced, high switching contrast $\tilde{I}_1$ is observed for all three duty cycles of the DS gratings. Moreover, the maximum of $\tilde{I}_1$ is found for $\lambda=1550\mathrm{~nm}$ with an average $\langle \tilde{I}_1 \rangle \sim3$. For the SD grating we observe a dramatic increase of the diffracted light intensity above the transition temperature due to initially weak contrast in the insulating state. Most notably, the measured $\tilde{I}_1$ for 1310\,nm and 1550\,nm exceed one order of magnitude. These values correspond to an enhancement of $\tilde{I}_1$ of DS gratings by factors of $\gtrsim 7$ and $\gtrsim8.5$. 
\begin{table}
   \centering
   \caption{$\tilde{I}_1$ for three DS gratings with different $DC$ and one SD grating.}
       \label{tab:1}
     \begin{tabular} {lcccccccc}
	Type & ~ & $DC$ &~& 980\,nm &~ & 1310\,nm & ~ & 1550\,nm \\ \hline \hline
	DS& ~ & 0.33 & ~ & $ 0.8\pm0.05 $ & ~ & $ 2.5\pm0.05 $ & ~ & $ 3.0\pm0.05$ \\  
	DS & ~ & 0.5 & ~ & $ 0.6\pm0.05$ & ~ & $2.1\pm0.05$ & ~ & $2.7\pm0.05$ \\ 
	DS & ~ & 0.66 & ~ & $0.8\pm0.05$ & ~ & $2.3\pm0.05$ & ~ & $3.2\pm0.05$ \\    
	SD & ~ & 0.5 & ~ & $7.3\pm0.05$ & ~ & $14.9\pm0.05$ & ~ & $22.8\pm0.05$ \\  \hline 
       \end{tabular}
\end{table}
\\

\begin{figure}
	\begin{center}
		\includegraphics[width=0.95\columnwidth]{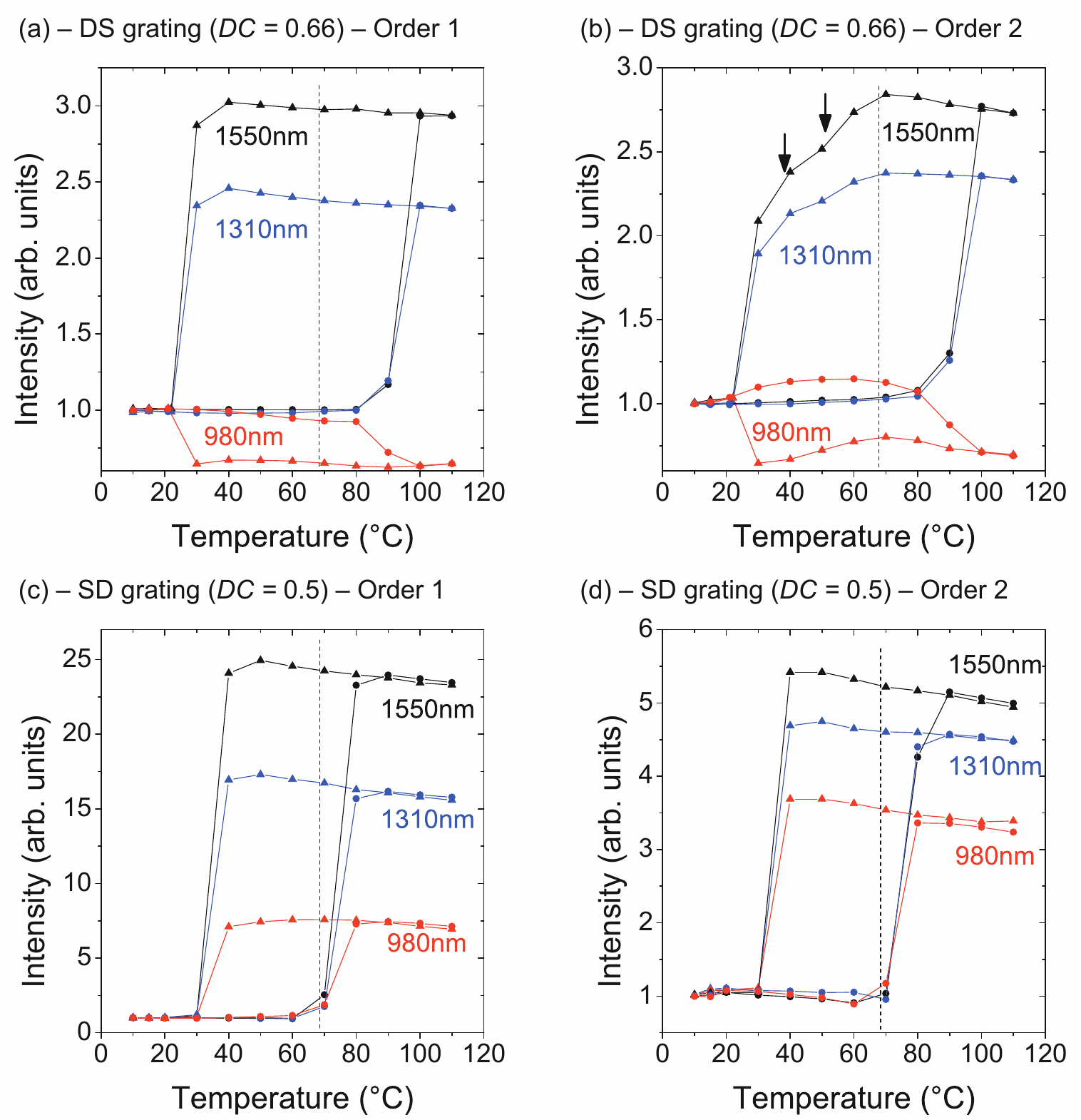}
		\caption{(Color online) Thermal hysteresis of $\tilde{I}_1$(\textit{left}) and $\tilde{I}_2$ (\textit{right}) for DS grating with $DC=0.66$ (a+b) and a SD grating with $DC=0.5$ (c+d) for the three wavelength studied. Intensities are normalized to the insulting phase and  circles (triangles) correspond to increasing (decreasing) temperature scans. The transition temperature of bulk crystals $(T_C=68\mathrm{~^oC})$ is marked by dashed lines.}
		\label{fig:3}
	\end{center}
\end{figure}

A well known effect in nanoscopic $\mathrm{VO_2}$ is a pronounced thermal hysteresis \cite{Roach:71,Lopez:01} which allows the implementation of bistable electronic and optical elements e.g. for memory and storage applications \cite{Driscoll:09,Xie:11}. We studied this effect for our ion beam synthesized $\mathrm{VO_2}$ nanothermochromic gratings and find wide hysteresis loops for both directly synthesized and selectively deactivated gratings. The obtained temperature dependencies of $\tilde{I}_1$ and $\tilde{I}_2$ for a DS grating ($DC=0.66$) and a SD grating ($DC= 0.5$) are plotted for increasing (circles) and decreasing (triangles) temperature and all three wavelengths studied in Fig. \ref{fig:3} (a,b) and (c,d), respectively. In these plots the transition temperature of bulk $\mathrm{VO_2}$ is marked by the dashed vertical line. For all samples studied the pronounced hysteresis is asymmetric with respect to $T_C$. Most notably, supercooling of the MIT persists down close to room temperature to $T=30\mathrm{~^oC}$ and $T=40\mathrm{~^oC}$ for the DS and SD gratings respectively. We attribute this remarkably broad hysteresis to an inhibition of the MIT by local strain fields within the embedded clusters or the absence of nucleation sites in small, single-domain $\mathrm{VO_2}$ nanocrystals. This interpretation is fully consistent with recent experiments demonstrating a wide range strain tuning of the MIT critical temperature in single-domain,free-standing $\mathrm{VO_2}$ nanobeams\cite{Wei:09,Cao:09}.\\

For both diffraction orders of the DS grating [cf. Fig. \ref{fig:3} (a) and (b)] the hysteresis is "reversed" for $\lambda=980$\,nm. This reversal directly reflects the reduced dielectric contrast. A similar observation was also reported by Suh \textit{et al.} for hole arrays in ${\rm Ag/VO_2}$ and ${\rm Ag/VO_2}$ structures\cite{Suh:2006}. In contrast, for the DS structures, the refractive index contrast increases for all three wavelength, giving rise to a "regular" hysteresis also for $\lambda=980$\,nm. The slightly wider hysteresis loop for the DS compared to the SD gratings is attributed to slightly different morphology and size distribution, although nominally identical synthesis parameters have been used. In the hysteresis of $\tilde{I}_2$ of the DS grating we observe a gradual, step-like decrease of the signal during the cooling cycle, marked by arrows in Fig. \ref{fig:3} (b). This observation points towards different switching temperatures of different nanocluster sub-ensembles due to variations of the local strain fields. Such characteristic step-like MIT has been previously observed for example in $\mathrm{VO_2}$ nanowires \cite{Wu:06} could be also present in the system studied here.\\

In summary, we fabricated nanothermochromic $(\mathrm{VO_2}/\mathrm{SiO_2})$ nanocrystal composite diffraction gratings using site selective ion implantation following two different approaches. The diffraction gratings fabricated by this standard semiconductor fabrication technique exhibit giant switching ratios of more than a factor of 3 for DS and exceeding one order of magnitude for SD gratings. Our technique can be readily extended to realize other diffractive planar optical elements such as phase lenses or optical memory devices. In the thermal hysteresis of higher diffraction orders we resolve clear fingerprints of nanostructure related blocking effects of the MIT using a contact-free, optical technique. This in turn opens new directions to investigate the currently widely investigated\cite{Whittaker:11} impact on the nanoscopic structural, chemical and morphological properties on the MIT of $\mathrm{VO_2}$ for large ensembles. In particular the unique anomalous increase of the diffraction efficiency for the investigated system could shed new light on the underlying physical mechanisms in nanoscopic $\mathrm{VO_2}$.\\

This work was financially supported by DFG via the Cluster of Excellence {\it Nanosystems Initiative Munich} (NIM) and the Emmy Noether Program (KR 3790/2-1).


\begin{thebibliography}{[1]}

\bibitem{Morin:59} F. J. Morin, Phys. Rev. Lett. \textbf{3}, 34--36 (1959).

\bibitem{Verleur:68} H. W. Verleur, A. S. Barker, C. N. Berglund, Phys. Rev \textbf{172}, 788--798 (1968).

\bibitem{Liu:11} W.-T. Liu, J. Cao, W. Fan, Z. Hao, M. C. Martin, Y. R. Shen, J. Wu, F. Wang, Nano Lett. \textbf{11}, 466--470 (2011).

\bibitem{Cavalleri:04a} A. Cavalleri, T. Dekorsky, H. H. W. Chong, J. C. Kieffer, R. W. Schoenlein, Phys. Rev. B \textbf{70}, 161102 (2004).

\bibitem{Rini:05a} M. Rini, A. Cavalleri, R. W. Schoenlein, R. Lopez, L. C. Feldman, R. F. Haglund Jr., L. A. Boatner, T. E. Haynes, Opt. Lett. \textbf{30}, 558--560 (2005).

\bibitem{Yang:11} Z. Yang, C. Ko, S. Ramanathan, Annu. Rev. Mater. Res. \textbf{41} 337--367 (2011).

\bibitem{Lopez:04a} R. Lopez, L. A. Boatner, T. E. Haynes, R. F. Haglund Jr., L. C. Feldman, Appl. Phys. Lett. \textbf{85}, 1410--1412 (2004).

\bibitem{Lee:89} C. E. Lee, R. A. Atkins, W. N. Giler, H. F. Taylor, Appl. Opt. \textbf{28}, 4511--4512 (1989).

\bibitem{Briggs:10} R. M. Briggs, I. M.; Pryce, H. A. Atwater, Opt. Express \textbf{18}, 11192--11201 (2010).

\bibitem{Lopez:02a} R. Lopez, L. A. Boatner, T. E. Haynes, L. C. Feldman, R. F. Haglund Jr., J. Appl. Phys. \textbf{92}, 4031--4036 (2002).

\bibitem{Li:10} S.-Y. Li, G. A. Niklasson, and C. G. Granqvist, J. Appl. Phys. \textbf{108}, 063525 (2010).

\bibitem{Tazawa:98} M. Tazawa, P. Jin, S. Tanemura, Appl. Opt. \textbf{37}, 1858--1861 (1998).

\bibitem{Roach:71} R. W. Roach, Appl. Phys. Lett. \textbf{19}, 453--455 (1971).

\bibitem{Lopez:01} R. Lopez, L. A. Boatner, T. E. Haynes, R. F.  Haglund Jr., L. C. Feldman, Appl. Phys. Lett. \textbf{79}, 3161--3163 (2001).

\bibitem{Driscoll:09} T. Driscoll, H.-T. Kim, B.-G. Chae, B.-J. Kim, Y.-W. Lee, N. M. Jokerst, S. Palit, D. R. Smith, M. Di Ventra, D. N. Basov, Science \textbf{325}, 1518--1521 (2009).

\bibitem{Xie:11} R. Xie, C. T. Bui, B. Varghese, Q. Zhang, C. H. Sow, B. Li, J. T. L. Thong, Adv. Funct. Mater. \textbf{21}, 1602--1607 (2011).

\bibitem{Wei:09} J. Wei, Z. H. Wang, W. Chen, D. H. Cobden, Nat. Nanotechnol. \textbf{4}, 420--424 (2009).

\bibitem{Cao:09} J. Cao, E. Ertekin, V. Srinivasan, W. Fan, S. Huang, H. Zheng, J. W. L. Yim, D. R. Khanal, D. F. Ogletree, J. C. Grossman, J. Wu, Nat. Nanotechnol. \textbf{4}, 732--737 (2009).

\bibitem{Suh:2006} J. Y. Suh, E. U. Donev, R. Lopez, L. C. Feldman, and R. F. Haglund Jr., Appl. Phys. Lett. \textbf{88}, 133115 (2006)

\bibitem{Wu:06} J. Wu, Q. Gu, B. S. Guiton, N. P. de Leon, L. Ouyang, H. Park, {Nano Lett.} \textbf{6}, 2313--2317 (2006).


\bibitem{Whittaker:11} L. Whittaker, C. J. Patridge, S. Banerjee, J. Phys. Chem. Lett. \textbf{2}, 745--758 (2011).



















\end{thebibliography}
\end{document}